\documentclass[a4paper,numberedheadings]{aip-cp}
\usepackage[utf8]{inputenc}
\usepackage[T1]{fontenc}
\usepackage[english]{babel}
\usepackage{graphicx}
\graphicspath{{Images/}}
\usepackage{url}
\usepackage{amsfonts}
\usepackage{amssymb}
\usepackage{hyperref}
\usepackage{geometry}
\usepackage{slashed}
\usepackage{bbm} 
\usepackage{soul}

\usepackage{tabularx}
\usepackage{color}

\newcommand{\e}{{\text{e}}}

\renewcommand{\i}{{\text{i}}}
\renewcommand{\Im}{{\text{Im}}}

\title{Semiclassical fermion pair creation in de Sitter spacetime}
\date{1 June 2015}
\author[ICRA,Rome,Nice]{Clément Stahl}
\eaddress{clement.stahl@icranet.org}
\author[ICRA,Rome,Nice]{Strobel Eckhard}
\eaddress{eckhard.strobel@irap-phd.eu}
\affil[ICRA]{ICRANet, Piazzale della Repubblica 10, 65122 Pescara, Italy}
\affil[Rome]{Dipartimento di Fisica, Universit\`a di Roma "La Sapienza", Piazzale Aldo Moro 5, 00185 Rome, Italy}
\affil[Nice]{Universit\'e de Nice Sophia Antipolis, 28 Avenue de Valrose, 06103 Nice Cedex 2, France}

\begin{document}

\maketitle

\begin{abstract}
We present a method to semiclassically compute the pair creation rate of bosons and fermions in de Sitter spacetime. The results in the bosonic case agree with the ones in the literature. We find that for the constant electric field the fermionic and bosonic pair creation rate are the same. This analogy of bosons and fermions in the semiclassical limit is known from several flat spacetime examples.
\end{abstract}

\section{Introduction}
Quantum field theory (QFT) in curved spacetime is one way of merging Einstein's theory of gravitation and QFT in the usual Minkowski spacetime within a self consistent framework  \cite{red book,birell}. One of its central results is the discovery of particle creation in a time-dependent gravitational field  \cite{ptcl creation}. This mechanism is believed to generate the primordial cosmic inhomogeneities that serve as seeds for the observed large-scale structure of the universe.\\
A lot of studies of QFT in curved spacetime are investigating de Sitter (dS) spacetime. It has constant scalar curvature and is maximally symmetric for any given dimension, in the same way as Minkowski spacetime. In cosmology, dS space is used as a model for both the early stage of inflation (for reviews see \cite{inflation}) and the late stage of acceleration of the expansion (an introductory review is given in \cite{DE}). In addition to its relevance for cosmology, dS space may give hints to understand the quantum nature of spacetime \cite{Qst}. These facts motivate the investigation of physical processes in this  spacetime.\\
Originally, particle creation was studied under a strong field background originating from a time dependent vector potential \cite{flat st}. Since then, it has been explored in a more general set-up, where the effects of the gravitational field and the electric field are both contributing to the creation of pairs. The Schwinger mechanism in curved spacetime, more specifically, in dS space, has recently started to be studied in depth  \cite{koba,vilen,holo,bubble original,1+1original,Villalbaspin1/2,kim,bub,kim2,kim3}.\\
Better understanding Quantum electrodynamics (QED) in dS spacetime could be insightful for a couple of problems. It is an interesting framework for the study of false vacuum decay or bubble nucleation \cite{vilen, bubble original}. Constrains on magnetogenesis scenarios were put in \cite{koba} via the backreaction of the created pairs and its induced current. In some models of preheating, it might give clues on the open problem of baryogenesis. Via the AdS-CFT correspondence, it has also been  used as a playing field to test the ER=EPR conjecture \cite{holo}. Additionally it might help to better understand renormalization schemes in curved space-time \cite{red book, landete} and the relations of these schemes to each other.\\
The purpose of this paper is to investigate spin 1/2 pair production under the influence of an external electric field, in four-dimensional dS spacetime ($\text{dS}_4$). This problem has already been studied for bosons in two-dimensional dS spacetime ($\text{dS}_2$) \cite{vilen} and in $\text{dS}_4$ \cite{koba}. In \cite{Villalbaspin1/2,dS2fermion} spin 1/2 pair production in $\text{dS}_2$ was investigated but the 4D analog case is much less studied. We propose a derivation of the number of pairs created in the semiclassical limit following the ideas of \cite{Strobel}. We find that there is no difference between bosons and fermions in this limit.\\
The paper is organized as follows: in section \ref{basics} we investigate the Schwinger effect in $\text{dS}_4$. After describing the basic equations, we show how the Dirac equation becomes a coupled second order differential equation in the presence of both electric and gravitational terms. We review the derivation of the bosonic pair creation rate and calculate the fermionic one with the help of a semiclassical expansion for general electric fields in dS spacetime in section \ref{WKB}. We use this result in section \ref{sec:const} to compute the bosonic and fermionic pair creation rate of a constant electric field in dS spacetime. Finally we draw some conclusions in section \ref{ccl}.

\section{Dirac field in dS Space-time}
\label{basics}

We consider QED coupled to a Dirac fermion in $\text{dS}_4$. In order to study Schwinger pair production, we assume that the gravitational and electrical field are background fields and that the fermionic field is dynamic. The action is given by
\begin{align}
\label{action}
S=\int \text{d}^4 \text{x} \sqrt{-g} \mathcal{L}=\int \text{d}^4 \text{x} \sqrt{-g} \left[  -\frac{1}{\kappa}R +\frac{i}{2} (\bar{\psi} \gamma^{\mu} \nabla_{\mu} \psi -\nabla_{\mu} \bar{\psi}\gamma^{\mu}\psi) -m \bar{\psi} \psi  -\frac{1}{4} F_{\mu \nu}F^{\mu \nu} \right],
\end{align} 
where the field strength is defined in the usual way $F_{\mu \nu}\equiv \partial_{\mu} A_{\nu}-\partial_{\nu} A_{\mu}$. \\
The $\text{dS}_4$ spacetime we want to study is described by the metric
\begin{equation} 
\text{ds}^2=a(\eta)^2 (\text{d}\eta^2-\text{d}\textbf{x}^2),
\end{equation}
with signature (+,-,-,-). Here \(\eta\) is the conformal time which is parametrized by the Hubble factor in the following way
\begin{equation}
\eta= -\frac{1}{a(\eta)H},\hspace{1cm} a(\eta)^2 H \equiv \frac{\text{d}a(\eta)}{\text{d} \eta }, \hspace{1cm} (-\infty<\eta<0).
\end{equation}
For the description of spinors the tetrad field is used. It is related to the metric through the relation
\begin{align}
g_{\mu \nu}=e^a_{\mu} e^b_{\nu} \eta_{ab},
\end{align}
where $\eta_{ab}$ is the usual Minkowski metric. Throughout this paper, we use Greek indices for spacetime indices ($\mu,\nu=\eta,x,y,z$) and Latin indices for tetrad ones ($a,b=\eta,x,y,z$). Applying the tetrad formalism to $\text{dS}_4$, one gets
\begin{equation}
 \label{tetrad} e_{\mu}^a=
\begin{pmatrix}
  a(\eta) & 0 & 0 &0  \\
  0 & a(\eta) & 0 & 0 \\
  0 & 0 & a(\eta) & 0 \\
  0 & 0 & 0 & a(\eta)
 \end{pmatrix}.
  \end{equation}
  The covariant derivative for fermion fields is defined as
\begin{align}
  \nabla_{\mu} \equiv \hbar\partial_{\mu} +ie A_{\mu}(x) -\frac{\i}{4}\omega^{ab}_\mu\sigma_{ab},
\end{align}
with the commutator of the gamma matrices 
\begin{align}
 \sigma_{ab}&\equiv\frac{\i}{2}[\gamma_a,\gamma_b]
\end{align}
and the spin connection are defined as
\begin{align}
\begin{split}
\omega_{\mu}^{ab}&\equiv \frac{1}{4}\left[e^{b\alpha}(x) \partial_{\mu}e^a_{\alpha}(x)-e^{a\alpha}(x) \partial_{\mu}e^b_{\alpha}(x)+ e^{a \alpha}(x) \partial_{\alpha}e^b_{\mu}(x)-e^{b\alpha}(x) \partial_{\alpha} e^{a \mu}(x)\right.\\ & \hspace{2cm}\left.+e^{b\nu}(x)e^{a \lambda}(x)e_{c \mu}(x)\partial_{\lambda} e^c_{\nu}(x)-e^{a\nu}(x)e^{b \lambda}(x)e_{c \mu}(x)\partial_{\lambda} e^c_{\nu}(x) \right]. \label{eq:spinconnection}
\end{split}
\end{align}
The non-zero components of the spin connection (\ref{eq:spinconnection}) can be shown to be 
  \begin{align}
 \label{connection}
 \omega_{1}^{01}=\omega_{2}^{02}=\omega_{3}^{30}=-\omega_{1}^{10}=-\omega_{2}^{20}=\omega_{3}^{30}=\frac{a'(\eta)}{2a(\eta)},
 \end{align}
where prime denotes derivative with respect to conformal time \(\eta\).\\
The gamma matrices $\gamma^{\mu}$ are related to the gamma matrices in the tangent flat space $\underline{\gamma^a}$ viz. 
\begin{align}
\underline{\gamma^a} \equiv \gamma_{\mu} e^a_{\mu}. 
\end{align} 
We will work with the Dirac representation of the gamma matrices, \emph{i.e.}
\begin{align}
 \gamma^{j}=\begin{pmatrix}
             0 &\sigma^j\\
             -\sigma^j & 0
            \end{pmatrix},
&&
  \gamma^{0}=\begin{pmatrix}
             I_2 &0\\
             0 & -I_2
            \end{pmatrix},
&& \text{ where }&&
   \sigma^x=\begin{pmatrix}
             0 & 1\\
             1 & 0
            \end{pmatrix},
&&
   \sigma^y=\begin{pmatrix}
             0 & -\i\\
             \i & 0
            \end{pmatrix},
            &&
   \sigma^z=\begin{pmatrix}
             1 & 0\\
             0 & -1
            \end{pmatrix}.
\end{align}
Varying the action with respect to the spinor field gives the Dirac equation
 \begin{equation}
\left(i \gamma^{\mu} \nabla_{\mu} -m \right) \psi(\textbf{x}, \eta) =0.
 \end{equation}
Using Eqs.~(\ref{tetrad}) and (\ref{connection}), this equation becomes
  \begin{equation}
  \label{eq:diracavantcahnge}
\left\{i \left(\gamma^{\mu} \hbar\partial_{\mu}+\frac{3}{2} a H \underline{\gamma^0} + i e A_{\mu} \gamma^{\mu} \right) -m \right\} \psi(\textbf{x}, \eta)=0.
 \end{equation}
One now considers the auxiliary field $ \Psi(\textbf{x}, \eta) = a^{3/2}(\eta) \psi(\textbf{x}, \eta)$ which can be thought of as the equivalent of the Mukhanov-Sasaki variable in inflation models \cite{red book}. With this substitution the Dirac equation takes the form
\begin{equation}
\left\{\gamma^{\mu}  ( i \hbar\partial_{\mu} - e A_{\mu})  - m \right\} \Psi(\textbf{x}, \eta)=0. \label{eq:Dirac}
 \end{equation}
 We will also decompose this field in momentum modes according to
\begin{align}
 \Psi(\textbf{x},\eta)\sim\e^{\frac{\i}{\hbar}\textbf{k}.\textbf{x}}\psi_{\textbf{k}}.\label{TF}
\end{align}
To solve the Dirac equation, for the purpose of the calculation of the pair creation rate, it is often useful to use the squared version of the Dirac equation because of its similarities to the Klein-Gordon equation, see e.g. \cite{Strobel,Dumlu2011}. The squared Dirac equation can be found using
 \begin{align}
  \Psi(\textbf{x},\eta)&=\gamma^{\mu}\left[(i \partial_{\mu}-e A_{\mu}(\eta)) +m a(\eta)\right]\phi(\textbf{x},\eta) \label{eq:squared}
 \end{align}
 with
 \begin{align}
 \phi_{\textbf{k}}(\eta)&=\begin{pmatrix}\phi_1(\eta)\\  \phi_2(\eta) \end{pmatrix},&&  \phi_i(\eta)=\begin{pmatrix} \phi_i^+(\eta)\\  \phi_i^-(\eta)\end{pmatrix}.
 \end{align}  
in the Dirac equation. In the previous equation, $ \phi_{\textbf{k}}(\eta)$ is the Fourier transform (in the sense of (\ref{TF})) of $\phi(\textbf{x},\eta)$. Here we consider a background vector potential for the electromagnetic sector such that
\begin{equation}
A_{\mu}\equiv A(\eta)\delta^{z}_{\mu}.
\end{equation}
For such fields the squared Dirac equation takes the form 
\begin{align}
&\left(\hbar^2 \partial_{\eta}^2+\omega_{\textbf{k}}(\eta)^2-i\hbar ma'(\eta) \right)\phi_1^{\pm} \pm i\hbar e A'(\eta) \phi_2^{\pm}(\eta)=0, \\
&\left(\hbar^2 \partial_{\eta}^2+\omega_{\textbf{k}}(\eta)^2+i\hbar ma'(\eta) \right)\phi_2^{\pm} \pm i\hbar e A'(\eta) \phi_1^{\pm}(\eta)=0,   \label{to solve}
\end{align}
where the effective pulsation and the kinetical momentum are defined as
\begin{align}
& \omega_{\textbf{k}}(\eta)^2\equiv p_z(\eta)^2+k_{\perp}^2+m^2 a(\eta)^2, \label{eq:omega}
& p_z(\eta)\equiv k_z+e A(\eta),
&& k_\perp^2\equiv k_x^2+k_y^2.
\end{align}
This equation can be compared to the equivalent bosonic  problem. The equation of motion derived from the Klein-Gordon equation is (see Eq~(2.13) of \cite{koba})
\begin{equation}
\label{boson to solve}
\left(\hbar^2 \partial_{\eta}^2+\omega_{\textbf{k},\text{boson}}(\eta)^2 \right) q_{\textbf{k}} =0,
\end{equation}
with
\begin{align}
\omega_{\textbf{k}. \text{boson}}(\eta)^2=\omega_{\textbf{k}}(\eta)^2-\frac{2}{\eta^2}. \label{eq:omegaboson}
\end{align}
The equation of the bosonic problem (\ref{to solve}) can be understood as a harmonic oscillator with a time dependent pulsation. The two other terms in Eq.~(\ref{to solve}) are new and due to the fermionic nature of the particles considered. On the one hand, the mass term was already derived \emph{e.g}. in \cite{landete} where no background electric field was considered. On the other hand, the electric term is present for instance in \cite{kluger} in flat spacetime. Such that this equation is a generalization of the Dirac equation in curved spacetime with background electric and gravitational field. \\
The squared Dirac equation is also analogous to the Dirac equation for two-component fields in flat spacetime. In \cite{Strobel} a method was used to semiclassically compute the pair creation rate for these fields. It is possible to use the same method for the case studied here. Instead of looking for a solution of the squared Dirac-equation (\ref{to solve}) we will however use the ansatz 
\begin{align}
\label{newansatz}
 \psi_{\vec{k},\uparrow}(\eta)=\begin{pmatrix}
                      -\i k_\perp \, \psi_2^+(\eta)\\
                       (k_x+\i k_y)\, \psi_2^+(\eta)\\
                      -\i k_\perp \, \psi_1^+(\eta)\\
                      -(k_x+\i k_y)\, \psi_1^+(\eta)\\
                     \end{pmatrix}, &&
   \psi_{\vec{k},\downarrow}(\eta)=\begin{pmatrix}
                      (k_x-\i k_y)\, \psi_2^-(\eta)\\
                      \i k_\perp \, \psi_2^-(\eta)\\
                      -(k_x-\i k_y) \, \psi_1^-(\eta)\\
                      \i k_\perp\, \psi_1^-(\eta)\\
                     \end{pmatrix} .                  
\end{align}
This ansatz can be derived by finding the solution of the squared equation analogous to \cite{Strobel} and then use (\ref{eq:squared}) to construct a solution for the Dirac equation (\ref{eq:Dirac}).
Observe that \(\psi_{\vec{k},\uparrow}(\eta)\) and \(\psi_{\vec{k},\downarrow}(\eta)\) are independent since
\begin{align}
 \psi_{\textbf{k},\uparrow}(\eta)^\dagger\cdot \psi_{\textbf{k},\downarrow}(\eta)=0.
\end{align}
Putting (\ref{newansatz}) in the Dirac equation (\ref{eq:Dirac}) leads to the equations
\begin{align}
 \i\hbar\, \psi_1^{'\pm}(\eta)+ m a(\eta)\, \psi_1^\pm(\eta)\pm (p_z(\eta)+\i k_\perp)\,\psi_2^\pm(\eta)=0,\label{to solve 1}\\
 \i\hbar\, \psi_2^{'\pm}(\eta)- m a(\eta)\, \psi_2^\pm(\eta)\pm (p_z(\eta)-\i k_\perp)\,\psi_1^\pm(\eta)=0,\label{to solve 2}
\end{align}
 which we will solve in section \ref{WKB}.
\section{Semiclassical number of pairs in dS spacetime}
\label{WKB}
The semiclassical expansion is considering cases where a vacuum state for the produced particles exits in the asymptotic future. That is true if the background fields are evolving slowly. To quantify the slow varying background more precisely, we introduce a dimensionless slowness parameter $T$ by replacing the scale factor $a(\eta)$ by a family of functions $a_T(\eta) \equiv a(\eta/T)$. Doing that, in the limit of infinitely slow varying backgrounds, $T \rightarrow \infty$, the derivatives of $a(\eta)$ will tend to zero. Orders of \(T\) are usually called adiabatic orders \cite{red book} but it can be noticed that the only place where \(T\) is involved is in the derivative $\partial_{\eta}$ of the Dirac equation so it is possible to "formally" pose  $T=1/\hbar$ and expand in power of $\hbar$.\\
In this section, we will first review the calculation of the number of particles in the bosonic case, which can be carried out with flat spacetime techniques and then present the fermionic case. The strategy will be the following:
\begin{itemize}
\item Reformulate the equation of motion in terms of an equation for the mode functions \(\alpha(\eta)\), \(\beta(\eta)\).
\item Perform a multiple integral iteration to compute $|\beta|^2$.
\item Calculate the integrals with a semiclassical saddle point approximation to derive the number of created pairs for each momentum mode \(\mathbf{k}\).
\end{itemize}
\subsection{Equations for the mode functions}
In this section we derive the equations for the mode functions from the Klein-Gordon and Dirac equation respectively. The aim is to construct equations in which \(\alpha'(\eta)\) depends only on \(\beta(\eta)\) and vice versa, in order to perform the multiple integral iteration of the next section.
\subsubsection{Bosonic case}
To compute the semiclassical pair creation rate we start from (\ref{boson to solve}).
The form of these equations is the same as in flat spacetime, the only difference being the specific time dependence of the fields. We will shortly review the well know techniques of flat spacetime (see e.g.~\cite{Dumlu2011}) applied to our field configurations. One can use the ansatz (which is inspired by a WKB expansion)
\begin{align}
  q_{\textbf{k}}(\eta)= &\frac{ \alpha(\eta)}{\sqrt{\omega_{\textbf{k}}(\eta)}}e^{-\frac{\i}{2} K_0(\eta) }+\frac{\beta(\eta)}{\sqrt{\omega_{\textbf{k}}(\eta)}}e^{\frac{\i}{2} K_0(\eta)},\label{eq:WKB-ansatzboson}\\
 q'_{\textbf{k}}(\eta)=&-\frac{\i\omega_{\textbf{k}}'(\eta)}{\hbar}\left[\frac{ \alpha(\eta)}{\sqrt{\omega_{\textbf{k}}(\eta)}}\e^{-\frac{\i}{2}K_0(\eta)}-\frac{\beta(\eta)}{\sqrt{\omega_{\textbf{k}}(\eta)}}\e^{\frac{\i}{2}K_0(\eta)}\right]\label{eq:WKB-ansatz2boson},
\end{align}
where
\begin{align}
K_0(\eta)=\frac{2}{\hbar} \int_{-\infty}^{\eta} \omega_{\textbf{k}}(\tau) d \tau, \label{eq:K}
\end{align}
where \(\alpha(\eta)\), \(\beta(\eta)\) are the mode functions.\\
The momentum spectrum of the pair creation rate is defined as  
\begin{align}
n_{\textbf{k}} \equiv \lim_{\eta\rightarrow 0} \left|\beta(\eta)\right|^2, \label{eq:trans}
\end{align}
with the boundary conditions
\begin{align}
 \beta(-\infty)=0, && \alpha(-\infty)=1. \label{eq:boundarycond}
\end{align}
Using this, together with the ansatz (\ref{eq:WKB-ansatzboson})-(\ref{eq:WKB-ansatz2boson}) in the Klein-Gordon equation (\ref{boson to solve}) we find that the mode functions are connected through coupled differential equations
\begin{align}
 \alpha'(\eta)&=\frac{\omega_{\textbf{k}}'(\eta)}{2\omega_{\textbf{k}}(\eta)}\e^{ i K_0(\eta)}\beta(\eta),\label{eq:alphaboson}\\
 \beta'(\eta)&=\frac{\omega_{\textbf{k}}'(\eta)}{2\omega_{\textbf{k}}(\eta)}\e^{-i K_0(\eta)}\alpha(\eta)\label{eq:betaboson}.
\end{align}

\subsubsection{Fermionic case}
To derive analogous equations in the fermionic case we start from the equations (\ref{to solve 1}) and (\ref{to solve 2}). 
The initial idea of this work comes from the similarities between this equations and Eq.~(11-12) of \cite{Strobel}.  For the semiclassical treatment one makes the following ansatz  
\begin{align}\psi^{\pm}_{1}(\eta)&=\frac{C_\pm}{\sqrt{\omega_{\textbf{k}}(\eta)}}\frac{\sqrt{p(\eta)}}{\sqrt{p_z(\eta)-\i k_\perp}}\left(\alpha^\pm(\eta)[\omega_{\textbf{k}}(\eta)-m a(\eta)]{\e^{-\frac{\i}{2}K(\eta)}} +\beta^\pm(\eta)[\omega_{\textbf{k}}(\eta)+m a(\eta)]{\e^{\frac{\i}{2}K(\eta)}}\right),\label{eq:ansatz1} \\
 \psi^{\pm}_{2}(\eta)&=\frac{\mp C_\pm}{\sqrt{\omega_{\textbf{k}}(\eta)}}\frac{\sqrt{p(\eta)}}{\sqrt{p_z(\eta)+\i k_\perp}}\left(\alpha^\pm(\eta)[\omega_{\textbf{k}}(\eta)+m a(\eta)]{\e^{-\frac{\i}{2}K(\eta)}} +\beta^\pm(\eta)[\omega_{\textbf{k}}(\eta)-m a(\eta)]{\e^{\frac{\i}{2}K(\eta)}}\right), \label{eq:ansatz2}
\end{align}
with the integrals
 \begin{align}
 K(\eta)& \equiv K_0(\eta)+K_1(\eta),\label{eq:K_s}\\
 K_1(\eta)& \equiv k_\perp\int_{-\infty}^{\eta} \frac{m a(\tau)p_z'(\tau)}{\omega_{\textbf{k}}(\tau)p(\tau)^2}d\tau \label{eq:K_xy}.
 \end{align}
 Using the ansatz (\ref{eq:ansatz1})-(\ref{eq:ansatz2}) in (\ref{to solve 1}) and (\ref{to solve 2}), we find that the mode functions are connected through coupled differential equations
\begin{align}
 \alpha'^\pm(\eta)&=-\frac{\omega_{\textbf{k}}'(\eta)}{2 \omega_{\textbf{k}}(\eta)}G_{\alpha}(\eta) \e^{\i K(\eta)}\beta^\pm(\eta),\label{eq:alphaferm}\\
 \beta'^\pm(\eta)&=\frac{\omega_{\textbf{k}}'(\eta)}{2 \omega_{\textbf{k}}(\eta)}G_{\beta}(\eta) \e^{-\i K(\eta)}\alpha^\pm(\eta).\label{eq:betaferm}
\end{align}
with
\begin{align}
& G_{\alpha}=\frac{ma(\eta)}{p(\eta)}-\frac{\omega_{\textbf{k}}(\eta)ma'(\eta)-\i k_\perp p_z'(\eta)}{p(\eta)\omega_{\textbf{k}}'(\eta)}, \\
&G_{\beta}=\frac{ma(\eta)}{p(\eta)}-\frac{\omega_{\textbf{k}}(\eta)ma'(\eta)+\i k_\perp p_z'(\eta)}{p(\eta)\omega_{\textbf{k}}'(\eta)},
\end{align}
representing the fermionic corrections to the analog bosonic case.
\subsection{Multiple integral iteration}
In this section we will perform the multiple integral iteration for the fermionic case. Because of the similar form of the equations (\ref{eq:alphaboson})-(\ref{eq:betaboson}) and (\ref{eq:alphaferm})-(\ref{eq:betaferm}) the results for the scalar case can be derived from the ones obtained below by setting \(G_\alpha(\eta)=G_\beta(\eta)=1\) and \(K(\eta)\rightarrow K_0(\eta)\).\\
By iteratively using Eqs.~(\ref{eq:alphaferm}) and (\ref{eq:betaferm}) and the boundary conditions (\ref{eq:boundarycond}) one finds
 \begin{equation}
 \begin{split}
& \beta^{\pm}(0)=\sum_{m=0}^\infty \int_{-\infty}^\infty d\eta_0\frac{\omega_k'(\eta_0)}{2\omega_k(\eta_0)}G_{\beta}(\eta_0)\e^{-iK(\eta_0)} \\
& \times \prod_{n=1}^m \int_{-\infty}^{\eta_{n-1}} d\tau_n\frac{\omega_k'(\tau_n)}{2\omega_k(\tau_n)}G_{\alpha}(\tau_n)\e^{iK(\tau_n)} \int_{-\infty}^{\tau_n}d\eta_n\frac{\omega_k'(\eta_n)}{2\omega_k(\eta_n)} G_{\beta}(\eta_n)\e^{-iK(\eta_n)}. \\\label{eq:multiint}
 \end{split}
\end{equation}
As described in \cite{Berry1982} these integrals are dominated by the classical turning points given by
\begin{align}
\label{TP}
 \omega_k(\eta_p^\pm)=0.
\end{align}
It can be shown (see \cite{Strobel,Dumlu2011,Berry1982}\footnote{The detailed intermediate steps can be found in Eqs.~(32)-(38) of \cite{Strobel}. Observe that in the current paper the integration contour is closed in the lower imaginary half plane because of opposite convention for the phases in (\ref{eq:multiint}).}) that for one pair of simple turning points the momentum spectrum of the pair creation rate (\ref{eq:trans}) in a semiclassical saddlepoint approximation is equal to
\begin{align}
 n^{\text{fermion}}_{\textbf{k}}=\left|\e^{-iK(\eta_p^{-})}\right|^2 \label{eq:MomentumSpectrumConst}.
\end{align}
For the bosonic case we can find
\begin{align}
 n_{\textbf{k}}^{\text{scalar}}=\left|\e^{-i K_0(\eta_p^{-})}\right|^2 .
\end{align}
by using the substitution discussed above. We thus find  that analogous to the case of two-component fields in flat space-time the difference between fermions and bosons is a factor of the form \(\exp(K_1(\eta_p^-))\).
\section{Pair creation rate of a constant electric field in dS spacetime}
\label{sec:const}
In this section we derive the pair creation rate for a constant electric field, which is described in $\text{dS}_4$ by
\begin{equation}
A(\eta)=\frac{E}{H^2 \eta}. 
 \end{equation}
This is due to the fact that a co-moving observer, with a four-velocity $u^{\mu}$, would measure an electric field of
\begin{equation}
E_{\mu} = u^{\nu} F_{\mu \nu}= a E\delta^{z}_{\mu},
 \end{equation}
which leads to a constant field strength since $E_{\mu}E^{\mu}=E^2$.\\
For convenience we introduce
\begin{equation}
\mu^2 = \frac{e^2 E ^2}{H^4}+ \frac{m^2}{H^2}= \lambda^2 + \gamma^2,
\end{equation}
where $\lambda= \frac{eE}{H^2}$ represent the electric field divided by the Hubble rate as usually in cosmological spacetime and $\gamma= \frac{m}{H}$ is the mass term divided by the Hubble rate. These two quantities represent the electric and gravitational contribution respectively. We can now check in which limit the semiclassical approximation holds. It requires the rate of change of the background to be small in the asymptotic future, that is
\begin{align}
\left(\frac{\omega'_{\textbf{k}}(\eta)}{\omega_{\textbf{k}}^2(\eta)}\right)^2 \underset{\eta \rightarrow 0}{\sim} \mu^{-2} &&\text{ and }&& \left(\frac{\omega''_{\textbf{k}}(\eta)}{\omega_{\textbf{k}}^3(\eta)}\right) \underset{\eta \rightarrow 0}{\sim} 2\mu^{-2}\end{align}
being small, we see that this is the case when $\mu \gg 1$. Comparing to the bosonic case we see that in the limit $\mu \gg 1$ the "-2" in (\ref{eq:omegaboson}) is negligible, so that the fermionic and bosonic pulsation are the same. To compute the pair creation rate we first have to compute the integrals \(K_0(\eta)\) and \(K_1(\eta)\) defined in (\ref{eq:K}) and (\ref{eq:K_xy}) respectively. The value of $\eta$ at the turning point (\ref{TP}) is found to be
\begin{equation}
\eta_p^{-}=\frac{-\lambda\frac{k_z}{k}-\i\sqrt{\gamma^2+\lambda^2\left(1-\left(\frac{k_z}{k}\right)^2\right)}}{k}.\label{eq:TP}
\end{equation}
The imaginary parts of $K_0(\eta_p^-)$ and $K_1(\eta_p^-)$ are the only ones contributing to (\ref{eq:MomentumSpectrumConst}). One can show that
\begin{align}
&\Im[K_0(\eta_p^-)]=-\pi\left(\mu-\frac{k_z}{k} \lambda\right)\theta(k_z \lambda),\\
& \Im[K_1(\eta_p^-)]=0,
\end{align}
Where \(\theta(x)\) is the Heaviside step function. It was introduced since the real part of the turning point (\ref{eq:TP}) has to be negative for the turning point to be inside of the closed contour which is needed in the approximation of (\ref{eq:multiint}). We thus find that only pairs with a momentum in \(z\)-direction which has the same sign as \(\lambda\) are produced in the semiclassical limit. This is what was called pair production in ``screening'' direction or ``downward'' tunneling in \cite{vilen}. The case of opposite sign, called ``upward'' tunneling is suppressed in the semiclassical limit.\\
$K_1(\eta_p^-)$ was the only difference between bosons and fermions and is not contributing to the number of pairs created. Thus we find that the number of pairs in the semiclassical limit for both bosons and fermions is given by 
\begin{equation}
\label{main}
n_{\textbf{k}} = \exp\left[-2\pi \left(\mu-\frac{k_z}{k} \lambda\right)\right] \theta(k_z \lambda).
\end{equation}
We will close this section by performing the flat spacetime limit and making the relation  with the bosonic case of \cite{koba} explicit.  The definition of the pair production rate is
\begin{equation}
\Gamma \equiv \frac{1}{(2 \pi)^3 V} \int d^3 \textbf{k} \,n_{\textbf{k}},
\end{equation} \\
where $ V= a(\eta)^4 d\eta$ is the unit four volume of the spacetime. As in \cite{koba}, an estimate for the moment when most of the particles are created can be found by analyzing when the adiabaticity is violated. This gives
\begin{equation}
\label{estimate}
\eta \sim -\frac{\mu}{k} .
\end{equation}
Using this the \(k\)-integral can be changed into a time integral. Going to spherical coordinates the $k_z$ integral can be performed.
Putting everything together, one finds
\begin{equation}
\Gamma= \frac{H^4}{(2\pi)^3} \frac{\mu^3}{|\lambda|} \left(\e^{2 \pi |\lambda|}-1\right) \e^{-2 \pi \mu}.
\end{equation}
We can compare this with Eq.~(2.33) of \cite{koba}, that we reproduce here
\begin{equation}
\Gamma = \frac{H^4}{(2 \pi)^3} \frac{(|\mu|^2+1/4)^{3/2}}{\sinh(2 |\mu| \pi)} \left\{\frac{H^2}{eE} \sinh\left(\frac{2 \pi eE}{H^2} \right) + 2 \pi e^{-2 |\mu| \pi} \right\}.
\end{equation}
We see that in the semiclassical limit $\mu \gg 1$ and $\lambda \gg 1 $ the expressions are equal. As in the bosonic case the physical number density \(n\) of produced pairs at the  time $\eta$ is given by
\begin{equation}
n=\frac{1}{a(\eta)^3} \int_{-\infty}^{\eta} d\tau a(\tau)^4 \Gamma= \frac{\Gamma}{3H}.
\end{equation}
The fact that it is constant shows that the dilution from the expansion of the universe is exactly compensated by the particles created from Schwinger and gravitational particle creation. Hence in the semiclassical limit, the population of fermions is always dominated by the particles created within a Hubble time.
The vacuum decay rate is defined for fermions as
\begin{equation}
\Upsilon=\log(1-|\beta_{\textbf{k}}|^2).
\end{equation}
The limit in which the Hubble parameter is negligible compared to the gravitational and electrical strength corresponds to the limit to flat spacetime. After some calculations which are analogous to the ones in \cite{koba} one can find the Minkowski limit by taking $H\rightarrow0$, which gives
\begin{equation}
\label{M4}
\begin{split}
& \lim_{H\rightarrow 0} \Gamma = \frac{(eE)^2}{(2 \pi)^3} \exp \left(-\frac{\pi m^2}{|eE|} \right), \\
& \lim_{H\rightarrow 0} \Upsilon =\sum_{i=1}^{\infty} \frac{1}{i^2} \frac{(eE)^2}{(2 \pi)^3} \exp \left(-\frac{i\pi m^2}{|eE|} \right).
\end{split}
\end{equation}
These are the familiar results for Schwinger pair production in Minkowski spacetime \cite{flat st}.
\section{Conclusions}
\label{ccl}
 In this work, we investigated the fermionic pair creation rate by the combination of an electric and a gravitational field in $\text{dS}_4$. We first presented the basic equations of our setup and derived the corresponding Dirac equation in section \ref{basics}. In section \ref{WKB}, we proposed a semiclassical approximation to compute the number of pairs produced. This approximation is based on a saddle point approximation of the integrals in (\ref{eq:multiint}). To make the comparison to the analog bosonic case easier, we first reviewed its computation within our formalism. Then we presented the computation of the number of fermions created for a constant electric field. It is the main result of this paper shown in (\ref{main}). The pair creation rate, in this limit, is the same in the fermionic case  as in the bosonic case. This equivalence between fermions and bosons in the semiclassical limit occurs also for one-component fields in flat spacetime when there is only one pair of turning points (see e.g. \cite{Strobel}). The limit to flat Minkowski spacetime is presented in (\ref{M4}) and agrees with the usual expression for the Schwinger effect.
 \\
Going beyond the semiclassical limit is possible but is out of the scope of this paper. To do this one can compute a more general quantity: the fermionic induced current. The current is proportional to the number of pairs created in the semiclassical limit but allows to explore a regime where the notion of adiabatic vacuum and of particles does not necessarily exist. It is a more precise way of describing the Schwinger effect because its definition does not depend on the time of creation of the pairs and the rough estimate (\ref{estimate}) can be avoided. A calculation of the current induced by an electric field in \(\text{dS}_2\) is performed in \cite{ourselves}. \\
Throughout this paper the electric and gravitational fields have been assumed to be external sources. One of the next steps could be to consider backreaction effects of the newly created particles to the external electric and gravitational fields. To do so, for the electromagnetic side, the electric current needs to be computed and be plugged into the equivalent of the Maxwell equation in curved spacetime. For the gravitational side, the density of particle needs to be plugged in the Friedman equation. Another direction could be to look for cosmological application \emph{e.g}. magnetogenesis, baryogenesis and the relation between dark energy and dark matter in the context of neutrino physics.

\section*{Acknowledgements} 
The authors thank Fernanda Gomes De Oliveira, Hendrik Ludwig, Pereira Jonas, Carlos Argüelles and She Sheng Xue for fruitful discussions. They are also grateful to all the people who were involved with the organization of the 2nd César Lattes Meeting in Rio de Janeiro. CS and ES are supported by the Erasmus Mundus Joint Doctorate Program by Grant Number 2012-1710 and 2013-1471 from the EACEA of the European Commission respectively.

\end{document}